\documentclass[twocolumn,showpacs,amsmath,amssymb]{revtex4-1}

\usepackage{setspace}
\usepackage{graphicx}
\usepackage{amsmath}
\usepackage{lipsum}
\usepackage{cases}
\usepackage{longtable}
\usepackage{afterpage}
\usepackage{rotating}
\usepackage{threeparttable}
\usepackage{booktabs}
\usepackage{multirow}
\usepackage{subfigure}
\usepackage{array}
\usepackage{color}
\usepackage{float}
\usepackage[section]{placeins}
\usepackage{longtable}

\usepackage{bm}
\usepackage{lipsum}
\usepackage{appendix}
\usepackage{soul}
\usepackage[colorlinks=true,linkcolor=red,citecolor=blue,urlcolor=red,]{hyperref}
\begin{document}

\renewcommand\arraystretch{1}
\title{Relativistic coupled-cluster calculation of hyperfine-structure constants of $^{229}$Th$^{3+}$ and evaluation of the electromagnetic nuclear moments of $^{229}$Th}
\author{Fei-Chen Li,$^{1,3}$ Hao-Xue Qiao,$^{1,*}$ Yong-Bo Tang,$^{2,*}$ and Ting-Yun Shi$^{3}$}

\affiliation {School of Physics and Technology, Wuhan University, Wuhan, 430072, China}
\affiliation {Physics Teaching and Experiment Center, Shenzhen Technology University, Shenzhen, 518118, China}
\affiliation {State Key Laboratory of Magnetic Resonance and Atomic and Molecular Physics, Wuhan
Institute of Physics and Mathematics, Innovation Academy for Precision Measurement Science and Technology, Chinese Academy of Sciences, Wuhan, 430071, China}
\email{qhx@whu.edu.cn,tangyongbo@sztu.edu.cn}
\date{\today}

\begin{abstract}
  $^{229}$Th is a promising candidate for developing nuclear optical clocks and searching for new physics beyond the standard model. For this purpose, it is important to have accurate knowledge of the nuclear properties of $^{229}$Th. In this work, we calculate hyperfine-structure constants for the lowest four states of $^{229}$Th$^{3+}$ using the relativistic coupled-cluster method based on the Gaussian basis set. The no-pair Dirac-Coulomb-Breit Hamiltonian with the lowest-order quantum electrodynamics (QED) correction is the starting point, and all linear and non-linear terms of single and double excitations are included in the coupled-cluster calculation. Combining the measured HFS constants [Phys. Rev. Lett. 106. 223001(2011)] and present atomic calculations, we extract the magnetic dipole moment, $\mu=0.359(9)$, and the electric quadrupole moment, $Q=2.95(7)$, of the $^{229}$Th nucleus. Our magnetic dipole moment is perfectly consistent with the recommended value from the all-order calculation by Safronova \textit{et. al.}[Phys.Rev.A 88, 060501(R) (2013)], but our electric quadrupole moment is smaller than their recommended value, by about $5\%$. A detailed analysis indicates that the non-linear terms of single and double excitations, not included in the all-order calculation, are crucial to produce a precise $Q$ value of $^{229}$Th. In addition, we also report the magnetic octupole hyperfine-structure constants and some important non-diagonal hyperfine transition matrix elements, which are required for further extraction of the magnetic octupole moment $\Omega$ of $^{229}$Th nucleus.

\end{abstract}

\pacs{ } \maketitle

\section{INTRODUCTION}

$^{229}$Th possesses an extremely low first excited isomeric state of only several eV~\cite{Kroger1976NuPhA,Beck2007PhRvL}, which opens the possibility to construct a high-precision nuclear optical clock~\cite{Peik2003EPL,Campbell2012PRL,Peik2021QS} and provides a strongly enhanced sensitivity for searching for new physics beyond the standard model~\cite{Safronova2018RMP,Flambaum2020PRA}. To achieve these attractive objects, accurate knowledge of the nuclear and electronic properties of $^{229}$Th, such as the nuclear moments, plays a fundamental role. The accurate nuclear moment values not only can help to grasp the hyperfine structures (HFS) of $^{229}$Th atoms and related ions, but also provide a benchmark tool to improve the nuclear model theory thereby helping to reliably predict the properties of its isomers such as the isomer transition rate and the nuclear moments~\cite{Minkov2019PRL}. The nuclear quadrupole moment of $^{229}$Th also can be used to assess the sensitivity of the isomeric transition to the possible variation of the fine-structure constant $\alpha$ and other fundamental constants~\cite{Flambaum2006PRL,Berengut2009PRL,Flambaum2020PRA}.

There have been some reported works on the magnetic dipole and electric quadrupole moment, $\mu$ and $Q$, of $^{229}$Th nucleus~\cite{Chasman1977RMP,Minkov2019PRL,Gerstenkorn1974JPF,Bemis_1988,Berengut2009PRL,Campbell2011prl,Safronova2013PRA2}. These results are summarized in Table~\ref{nuclear1}, and one can find obvious discrepancies between the predicted values from the nuclear theory~\cite{Minkov2019PRL,Chasman1977RMP} and the deduced values from atomic structure calculation combined with experimental measurements~\cite{Gerstenkorn1974JPF,Bemis_1988,Berengut2009PRL,Safronova2013PRA2}. For $\mu$, the previous nuclear theory predictions~\cite{Minkov2019PRL,Chasman1977RMP} are significantly above the experimentally deduced values~\cite{Gerstenkorn1974JPF,Safronova2013PRA2}, but for $Q$, the nuclear theoretical prediction~\cite{Minkov2019PRL} is lower than the experimentally deduced values~\cite{Gerstenkorn1974JPF,Bemis_1988,Berengut2009PRL,Safronova2013PRA2}. Among these results, the values extracted by Safronova \textit{et al.}~\cite{Safronova2013PRA2} using all-order calculations based on $^{229}$Th$^{3+}$ combined with precise hyperfine spectral measurement~\cite{Campbell2011prl} are considered the most reliable~\cite{2018MolPh}. In their calculations, all possible single and double excitations are iterated to all orders of perturbation theory, and a part of triple excitations are included perturbatively, thus it is termed as SDpT method. However, only the linear coupled-cluster terms of single- and double-excitations are included in the SDpT method, and we notice that the correlation effects represented by the nonlinear terms play a crucial role in the precise prediction for the HFS of Fr or Fr-like monovalent systems~\cite{Sahoo2015PRA,li2021}. Thus it is necessary and important to accurately assess the contribution of nonlinear terms in the HFS calculations of Fr-like $^{229}$Th$^{3+}$.

\begin{table}[ht]\small
  \begin{threeparttable}
  \newcommand{\RNum}[1]{\uppercase\expandafter{\romannumeral #1\relax}}
  \caption{The magnetic dipole moment $\mu$ ($\mu_{N}$) and electric quadrupole moment $Q$ ($e$b) of $^{229}$Th reported in previous studies. ``Theoretical prediction'' results are obtained by calculations based on various nuclear theoretical models. ``Experimentally deduced'' values are found from laser spectroscopy measurements in combination with atomic calculations.
  The $Q$ of the penultimate column was derived by Campbell \textit{et al.}~\cite{Campbell2011prl}, combining their measurements and the calculations by Berengut \textit{et al.}~\cite{Berengut2009PRL}. The $\mu$ and $Q$ of the last column were derived by Safronova \textit{et al.}~\cite{Safronova2013PRA2}, combining their calculations and the measurements by Campbell \textit{et al.}~\cite{Campbell2011prl}.}\label{nuclear1}
  \begin{ruledtabular}
  \begin {tabular}{ccccccccc}
   \multicolumn{1}{c}{\multirow{2}{*}{}}
   &\multicolumn{2}{c}{Theoretical prediction}
   &\multicolumn{1}{c}{}
   &\multicolumn{4}{c}{Experimentally deduced}\\
  \cline{2-3}\cline{5-8}

  \multicolumn{1}{c}{}
  &\multicolumn{1}{c}{Ref.~\cite{Minkov2019PRL}}
  &\multicolumn{1}{c}{Ref.~\cite{Chasman1977RMP}}
  &\multicolumn{1}{c}{}
  &\multicolumn{1}{c}{Ref.~\cite{Gerstenkorn1974JPF}}
  &\multicolumn{1}{c}{Ref.~\cite{Bemis_1988}}
  &\multicolumn{1}{c}{Ref.~\cite{Campbell2011prl}}
  &\multicolumn{1}{c}{Ref.~\cite{Safronova2013PRA2}}\\

  \hline
    $\mu$  &0.530$-$0.655 &0.54 &&0.46(4)  &       &        &0.360(7)   \\
    $Q$    &2.80          &     &      &4.3(9)   &3.15(3)&3.11(16)&3.11(6)    \\
  \end{tabular}
  \end{ruledtabular}
  \end{threeparttable}
  \end{table}

Theoretical investigation for the HFS of $^{229}$Th$^{3+}$ is also of much interest in some other aspects. It has been reported as the most promising system for the realization of a single-ion nuclear clock based on a virtual clock transition composed by a pair of stretched hyperfine states of both nuclear ground and isomeric manifolds~\cite{Campbell2012PRL}. Knowledge of the relevant hyperfine interaction matrix elements is helpful to extract the nuclear excitation energy~\cite{Beloy2014PRL}. Moreover, $^{229}$Th$^{3+}$ is also a possible system for atomic parity-nonconservation experiments~\cite{Roberts2013PRA}, and its magnetic dipole HFS constant would be a reliable probe for the estimate of theoretical accuracy of the related weak matrix elements. \\

In the present work, we calculate the HFS constants of the low-lying states of $^{229}$Th$^{3+}$ using the relativistic coupled-cluster method based on the Gaussian basis set. The correlation effects are investigated by the \textit{ab initio} methods at different levels including the Dirac-Fock approximation, low-order many-body perturbation theory, linearized and fully single- and double-excitation relativistic coupled-cluster method. The magnetic dipole and electric quadrupole moment, $\mu$ and $Q$, of $^{229}$Th nucleus are extracted by combining present recommended theoretical results with the available experimental values. The hyperfine interaction matrix elements required to extract the magnetic octupole moment are also presented. In section II, we provide a brief overview of the coupled-cluster method and the hyperfine structure theory. Numerical results and discussions are presented in section III. Finally, a summary is given in section IV.

\section{Theoretical method}
\subsection{A brief description of the relativistic coupled-cluster approach}
For Th$^{3+}$, being a monovalent atomic system, the exact wavefunction with a valence orbital $\upsilon$ can be formally expressed in the form
\begin{small}
\begin{equation}
  \label{eqcc1}
|\Psi_{\upsilon}\rangle= e^{S}|\Phi_{\upsilon}\rangle,
\end{equation}
\end{small}
where $|\Phi_{\upsilon}\rangle$ is the zeroth-order wavefunction obtained by the Dirac-Fock calculation, and the exponential cluster operator, $e^{S}$ = 1+$S$+$\frac{1}{2!}S^{2}$+$\dots$, represents the expansion of wave operators in the framework of the coupled-cluster theory. According to the number of particles, $n$, to be excited from the reference configuration, the cluster operator $S$ can be partitioned into
\begin{equation}
  S=S_{1}+S_{2}+\dots+S_{n}.
\end{equation}
In practice, the contributions from the triple and higher excitations are always expected to be relatively small and the computations are extremely time-consuming. Thus, only the singles ($S_{1}$) and doubles ($S_{2}$) excitations are considered in the present coupled-cluster calculation (CCSD), and Eq.(\ref{eqcc1}) can be written as
\begin{equation}
  \label{eqcc2}
  \begin{aligned}
  \left|\Psi_{v}\right\rangle_{\rm CCSD} = &\left\{1+S_{1}+S_{2}+\frac{1}{2!}\left(S_{1}^{2}+S_{2}^{2}+S_{1} S_{2}\right)\right.\\
  &\left.+\frac{1}{3!}\left(S_{1}^{3}+3 S_{1}^{2} S_{2}\right)+\frac{1}{24} S_{1}^{4}\right\}\left|\Phi_{v}\right\rangle.
  \end{aligned}
  \end{equation}
\\

Eq.(\ref{eqcc2}) is a nearly complete expression of the wavefunction within the single and double excitation approximation (SD), where the first three terms are linear terms and provide the majority of the contribution. A more convenient and simplified treatment is to retain only the linear terms, which are referred to as the linearized coupled-cluster (LCCSD) wavefunction,
  \begin{equation}
    \label{eqcc3}
    |\Psi_{\upsilon}\rangle_{\rm LCCSD}=(1+S_{1}+S_{2})|\Phi_{\upsilon}\rangle.
  \end{equation}
However, for some strongly correlated systems, the contribution from nonlinear terms may be important and need to be evaluated carefully, especially in the high precision calculations of some special properties.

The transition matrix elements of an operator $\hat{O}$ from state  $\left|\Psi_{w}\right\rangle$ to $\left|\Psi_{v}\right\rangle$ can be evaluated according to
\begin{equation}
  \label{eqcc4}
\begin{aligned}
\bar{O} &=\frac{\left\langle\Psi_{w}|\hat{O}| \Psi_{v}\right\rangle}{\left\langle\Psi_{w} \mid \Psi_{v}\right\rangle} \\
&=\frac{\left\langle\Phi_{w}\left|e^{S \dagger} \hat{O} e^{S}\right| \Phi_{v}\right\rangle}{\left\langle\Phi_{w}\left|e^{S \dagger} e^{S}\right| \Phi_{v}\right\rangle},
\end{aligned}
\end{equation}
where ${e^{S \dagger}}$ stands for the complex conjugate of $e^{S}$. Expanding the wavefunction $\Psi$ using Eq.(\ref{eqcc2}) or Eq.(\ref{eqcc3}), one would obtain the transition matrix element within the CCSD or LCCSD approximation, respectively. The specific calculation steps are described in detail in the previous work~\cite{li2021JPhB,li2021}. \\

\subsection{The hyperfine structure theory}
The hyperfine interaction Hamiltonian for a relativistic electron can be expressed as~\cite{Schwartz1955PR}
\begin{small}\begin{equation}
H_{\mathrm{HFI}}=\sum_{k}T_{k}^{e}\cdot T_{k}^{n},
\end{equation}\end{small}
where $T_{k}^{e}$ and $T_{k}^{n}$ are the spherical tensor operators with rank $k$ ($k>0$) in the electronic and nuclear coordinates, respectively.
 A matrix element of the HFI between the basis of the hyperfine states $|\alpha I,\gamma J; F M_{F}\rangle$, which coupled a nuclear eigenstate $|\alpha I,M_{I}\rangle$ and an atomic eigenstate $|\gamma J,M_{J}\rangle$ is

\begin{small}
  \begin{equation}\begin{aligned}
   &\left\langle \alpha^{\prime} I^{\prime},\gamma^{\prime} J^{\prime}; F^{\prime} M_{F}^{\prime}\left|H_{\mathrm{HFI}}\right| \alpha I,\gamma J; F M_{F}\right\rangle =\delta_{F^{\prime} F} \delta_{M_{F}^{\prime} M_{F}} \\
  &\times (-1)^{I+J+F} \sum_{k}\left\{\begin{array}{lll}
    F & J & I \\
    k & I & J^{\prime}
    \end{array}\right\}\left\langle\gamma^{\prime} J^{\prime}\left\|T_{k}^{e}\right\| \gamma J\right\rangle\left\langle \alpha^{\prime} I^{\prime}\left\|T_{k}^{n}\right\|\alpha I\right\rangle,
  \end{aligned}\end{equation}
\end{small}
where $F$ is the total angular momentum with $\textbf{F}$=$\textbf{I}$+$\textbf{J}$, $I$ is the nuclear spin, $J$ is the total electronic angular momentum, $\alpha$ and $\gamma$ encapsulate the remaining nuclear and electronic quantum numbers, respectively. Then the first-order correction $E_{F, J}^{(1)}$ of hyperfine interaction to the energy is defined as
\begin{small}
  \begin{equation}
    \label{e1}
    \begin{aligned}
    E_{F}^{(1)}=(-1)^{I+J+F}\sum_{k}\left\{\begin{array}{lll}
    F & J & I \\
    k & I & J
    \end{array}\right\}\left\langle\gamma J\left\|T_{k}^{e}\right\| \gamma J\right\rangle\left\langle \alpha I\left\|T_{k}^{n}\right\| \alpha I\right\rangle,
    \end{aligned}
    \end{equation}
  \end{small}

Restricted to $k\leq3$, $E_{F}^{(1)}$ can be parameterized in terms of the HFS constants $A$, $B$, and $C$. These HFS constants are expressed as follows, respectively

\begin{small}
\begin{equation}\begin{aligned}
A =\mu_{N}\frac{\mu_{I}}{I}\frac{\langle {\gamma J|}\big |{ T_{1}^{e}}\big |{|\gamma J}\rangle}{\sqrt{J(J+1)(2J+1)}},
\end{aligned}\end{equation}
\end{small}
\begin{small}
\begin{equation}\begin{aligned}
B =2Q \left [\frac{2J(2J-1)}{(2J+1)(2J+2)(2J+3)}\right ]^{1/2}{\langle {\gamma J|}\big |{ T_{2}^{e}}\big |{|\gamma J}\rangle},
\end{aligned}\end{equation}
\end{small}
and
\begin{small}
\begin{equation}\begin{aligned}
C =\Omega_{I} \left [\frac{J(2J-1)(J-1)}{(J+1)(J+2)(2J+1)(2J+3)}\right ]^{1/2}{\langle {\gamma J|}\big |{ T_{3}^{e}}\big |{|\gamma J}\rangle},
\end{aligned}\end{equation}
\end{small}
where $A$, $B$, and $C$ are magnetic dipole (M1), electric quadrupole (E2), and magnetic octupole (M3) hyperfine structure constants respectively. $\mu_{N}$ is the nuclear Bohr magneton, and the diagonal nuclear reduced matrix elements $\left\langle \alpha I\left\|T_{k}^{n}\right\|\alpha I\right\rangle$ in Eq.(\ref{e1}) are contained in the nuclear moments, $\mu_{I}$ ($k$=1), $Q$($k$=2), $\Omega_{I}$ ($k$=3), respectively.

The contributions from second-order hyperfine interactions are generally on the same order as the magnetic octupole contribution, thus we also take into account these terms here. In addition, due to the first excited nuclear state of $^{229}$Th has an anomalously small excitation energy, there should also be a small but observable contribution in second order correction that corresponds to the hyperfine mixing between the electronic states of the ground nuclear states and the electronic states of isomer nuclear states~\cite{Beloy2014PRL}.
Then the second-order correction $E_{F}^{(2)}$ of hyperfine interaction to the energy in $^{229}$Th is defined as
%\alpha^{\prime} I^{\prime}
\begin{small}
  \begin{equation}
    \begin{aligned}
    E_{F}^{(2)}&= \sum{\prime} \frac{1}{E_{\alpha I,\gamma J}-E_{\alpha^{\prime} I^{\prime},\gamma^{\prime} J^{\prime}}} \sum_{k_{1}, k_{2}}\left\{\begin{array}{lll}
    I & J & F \\
    J^{\prime} & I & k_{1}
    \end{array}\right\}\left\{\begin{array}{lll}
    I & J & F \\
    J^{\prime} & I & k_{2}
    \end{array}\right\} \\
    & \times\left\langle \alpha I\left\|T_{k_{1}}^{n}\right\| \alpha^{\prime} I^{\prime}\right\rangle\left\langle \alpha^{\prime} I^{\prime}\left\|T_{k_{2}}^{n}\right\| \alpha I\right\rangle\left\langle\gamma J\left\|T_{k_{1}}^{e}\right\| \gamma^{\prime} J^{\prime}\right\rangle\left\langle\gamma^{\prime} J^{\prime}\left\|T_{k_{2}}^{e}\right\|\gamma J \right\rangle.
    \end{aligned}
    \end{equation}
  \end{small}
The summation involves all possible excited nuclear states and electronic states, and $E_{\alpha I,\gamma J}$ includes both nuclear and electronic energies. $\left\langle \alpha I\left\|T_{k}^{n}\right\| \alpha^{\prime} I^{\prime}\right\rangle$ and $\left\langle\gamma J\left\|T_{k}^{e}\right\| \gamma^{\prime} J^{\prime}\right\rangle$ correspond to the reduced matrix elements of nuclear part and electronic parts, respectively.  The off-diagonal reduced matrix elements of nuclear part can be found in Ref.~\cite{Beloy2014PRL}. In the present work, we focus on M1 and E2 off-diagonal reduced matrix elements from the fine-structure splitting (electronic part), because their contributions dominate owing to small energy denominators.

The single particle reduced matrix elements of the operators $T_{1}^{e}$, $T_{2}^{e}$, and $T_{3}^{e}$ are given by
\begin{small}
\begin{eqnarray}\label{eq:88}
\langle{\kappa_{i}}&\|T_{1}^{e}\|&\kappa_{j}\rangle=-\langle-\kappa_{i}\|C^{(1)}\|\kappa_{j}\rangle
(\kappa_{i}+\kappa_{j})\notag\\
&\times&\int_{0}^{\infty}{dr\frac{P_{i}(r)Q_{j}(r)+P_{j}(r)Q_{i}(r)}{r^2}\times{F(r)}},
\end{eqnarray}
\end{small}
\begin{small}
%\end{widetext}
\begin{eqnarray}
\langle{\kappa_{i}}\|T_{2}^{e}\|\kappa_{j}\rangle&=&-\langle\kappa_{i}\|C^{(2)}\|\kappa_{j}\rangle\notag\\
&\times&\int_{0}^{\infty}{dr\frac{P_{i}(r)P_{j}(r)+Q_{j}(r)Q_{i}(r)}{r^3}},
\end{eqnarray}
\end{small}
and
\begin{small}
\begin{eqnarray}
\langle{\kappa_{i}}\|T_{3}^{e}\|\kappa_{j}\rangle&=&-\frac{1}{3}\langle-\kappa_{i}\|C^{(3)}\|\kappa_{j}\rangle
(\kappa_{i}+\kappa_{j})\notag\\
&\times&\int_{0}^{\infty}{dr\frac{P_{i}(r)Q_{j}(r)+P_{j}(r)Q_{i}(r)}{r^4}},
\end{eqnarray}
\end{small}
where the relativistic angular-momentum quantum number $\kappa=\ell(\ell+1)-j(j+1)-1/4$, $P$ and $Q$ are the large and small radial components of Dirac wavefunction, respectively.

In the present work, we employed a finite basis set composed of even-tempered Gaussian-type functions expressed as $G_{i}=\mathcal N_{i}r^{\ell+1}e^{-\alpha_{i}r^{2}}$, to expand the Dirac radial wavefunction $P$ and $Q$ as in Refs.~\cite{Chaudhuri1999pra,li2021JPhB,li2021}, where $\mathcal N_{i}$ is the normalization factor, and $\alpha_{i}=\alpha\beta^{i-1}$, where the two independent parameters $\alpha$ and $\beta$ are optimized separately for each orbital symmetries. Table~\ref{basis} lists the Gauss basis parameters. $N$ is the number of basis set for each symmetry. $N_{c}$ and $N_{v}$ represent the number of core and virtual orbitals, respectively.
 \begin{table}[H]
  \newcommand{\RNum}[1]{\uppercase\expandafter{\romannumeral #1\relax}}
  \caption{The parameters of the Gauss basis set. $N$ is the number of basis set for each symmetry. $N_{c}$ and $N_{v}$ represent
  the number of core orbitals and virtual orbitals, respectively. }\label{basis}
  \renewcommand\tabcolsep{2.0pt}
  \begin{ruledtabular}
  \begin {tabular}{lcccccccc}
    &$s$&$p$&$d$&$f$&$g$&$h$&$i$&$k$
    \\   \hline
    $\alpha\times10^{3}$ &3.3&3.0&4.6&13.5&12&22&23&24\\
    $\beta$ &1.95   &1.95   &1.695   &1.681   &1.95   &2.05  &2.15   &2.25\\
    $N$      &40     &38     &35     &32     &25    &20 &15    &10  \\
    $N_{c}$  &6      &5      &3      &1      &0     &0 &0     &0  \\
    $N_{v}$  &22     &22     &26     &24     &19    &16 &15    &10  \\
  \end{tabular}
  \end{ruledtabular}
  \end{table}

In our calculations, the no-pair Dirac Hamiltonian is set as the starting point. Breit interaction and the lower-order quantum electrodynamics (QED) radiative potential proposed by Flambaum and Ginges~\cite{Flambaum2005pra} are considered at the same foot as coulomb interaction. The Fermi nuclear distribution was employed to describe the Coulomb potential between electrons and the nucleus. All the core orbitals and virtual orbitals with energies of smaller than 10000 a.u. are included in the correlation calculations.

\section{RESULTS AND DISCUSSION}

\subsection{Energies}

\begin{table*}[ht]\small
\begin{threeparttable}
\newcommand{\RNum}[1]{\uppercase\expandafter{\romannumeral #1\relax}}
\caption{Zeroth-order DF Coulomb correction ${E^{(0)}}$, second-order MBPT correction ${E^{(2)}}$, linearized part of single ($S_{1}$) and double ($S_{2}$) excitations correction ${E^{(S_{1}+S_{2})}}$, the non-linear coupled-cluster terms corrections ${E^{\rm(NL)}}$, the corrections of Breit and QED effects to the eneriges for Th$^{3+}$ in cm$^{-1}$. ${E_{\rm MBPT(2)}}$=${E^{(0)}}$+Breit+${ E^{(2)}}$ , ${ E_{\rm LCCSD}}$=${E^{(0)}}$+ Breit+${E^{(S_{1}+S_{2})}}$ , ${E_{\rm CCSD}}$=${ E_{\rm LCCSD}}$+ ${E^{\rm (NL)}}$, ${ E_{\rm Final}}$=${E_{\rm CCSD}}$+${\rm QED}$, represent the results obtained within second-order MBPT, LCCSD, CCSD, and CCSD results with QED correction approximations, respectively. The experiment energies ${ E_{\rm Expt.}}$ with the uncertainty in parentheses are also listed for comparison.}\label{energy1}
\begin{ruledtabular}
\begin {tabular}{lccccccccccc}
  Level&${E^{(0)}}$&Breit&${E^{(2)}}$& ${E^{(S_{1}+S_{2})}}$ & ${E^{\rm (NL)}}$ &QED& ${ E_{\rm MBPT(2)}}$& ${ E_{\rm LCCSD}}$  & ${ E_{\rm CCSD}}$ &${ E_{\rm Final}}$ &${ E_{\rm Expt.}}$~\cite{NIST,actinide}\\ \hline
  $5f_{5/2}$&$-$206612  &$-$737   &$-$31348 &$-$25459 &754  &$-$204  &$-$238697&$-$232808&$-$232054&$-$232258&$-$231065(200) \\
  $5f_{7/2}$&$-$203185  &$-$869   &$-$29828 &$-$24413 &754  &$-$198  &$-$233882&$-$228467&$-$227714&$-$227911&$-$226740 \\
  $6d_{3/2}$&$-$211800  &$-$48    &$-$12980 &$-$11154 &441  &$-$89   &$-$224828&$-$223002&$-$222562&$-$222650&$-$221872 \\
  $6d_{5/2}$&$-$207687  &$-$129   &$-$12314 &$-$10923 &422  &$-$68   &$-$219055&$-$217663&$-$217241&$-$217308&$-$216579 \\
  $7s_{1/2}$&$-$200273  &$-$90    &$-$10931 &$-$9173  &378  &$-$182  &$-$211294&$-$209536&$-$209157&$-$208975&$-$207934 \\
  $7p_{1/2}$&$-$165094  &$-$167   &$-$7587  &$-$6746  &365  &$-$3    &$-$172848&$-$172007&$-$171643&$-$171645&$-$170826 \\
  $7p_{3/2}$&$-$153571  &$-$57    &$-$6042  &$-$5450  &319  &$-$7    &$-$159670&$-$159079&$-$158760&$-$158753&$-$158009 \\
\end{tabular}
\end{ruledtabular}
\end{threeparttable}
\end{table*}

\begin{table*}[ht]
\begin{threeparttable}
\newcommand{\RNum}[1]{\uppercase\expandafter{\romannumeral #1\relax}}
\caption{Comparison of our calculated ${ E_{\rm LCCSD}}$, ${ E_{\rm CCSD}}$, ${ E_{\rm Final}}$ transition energies with other available theoretical and experimental data in ${\rm cm^{-1}}$, the difference of theoretical and experimental data labeled by $\triangle$. The first row lists the absolute energy of the ground state, and the other rows list the excitation energies of other states with respect to the ground state. }\label{energy2}
\begin{ruledtabular}
\begin {tabular}{cccccccccccc}
\multicolumn{1}{c}{Level}
&\multicolumn{1}{c}{${ E_{\rm LCCSD}}$}
&\multicolumn{1}{c}{$\triangle$}
&\multicolumn{1}{c}{${ E_{\rm CCSD}}$}
&\multicolumn{1}{c}{$\triangle$}
&\multicolumn{1}{c}{${ E_{\rm Final}}$}
&\multicolumn{1}{c}{$\triangle$}
&\multicolumn{1}{c}{${ E_{\rm FSCCSD}}$~\cite{ELIAV2012CP}}
&\multicolumn{1}{c}{$\triangle$}
&\multicolumn{1}{c}{${ E_{\rm all-order}}$~\cite{Safronova2006PRA}}
&\multicolumn{1}{c}{$\triangle$}
&\multicolumn{1}{c}{${{E_{\rm Expt.}}}$~\cite{NIST,actinide}}\\
\hline
$5f_{5/2}$ &$-$232808 &$-$1743 &$-$232054 &$-$989 &$-$232258 &$-$1193 &$-$231957 &$-$892 &$-$230304 & 761 &$-$231065\\
$5f_{7/2}$  & 4341 &16      & 4340 &15     & 4346 & 21     & 4320 &$-$5   &4136 & $-$189  &4325 \\
$6d_{3/2}$  & 9806 &613     & 9492 &299    & 9608 &415     & 9416 &223    &8304 & $-$889  &9193 \\
$6d_{5/2}$  &15145 &659     &14812 &326    &14949 &463     &14738 &252    &13377&$-$1109  &14486 \\
$7s_{1/2}$  &23272 &141     &22896 &$-$235 &23282 &151     &22833 &$-$308 &22229&$-$902   &23131 \\
$7p_{1/2}$  &60800 &561     &60411 &172    &60612 &373     &60346 &109    &59213&$-$1026  &60239 \\
$7p_{3/2}$  &73729 &673     &73293 &237    &73504 &448     &73206 &150    &71932&$-$1124  &73056 \\
\end{tabular}
\end{ruledtabular}
\end{threeparttable}
\end{table*}

We carried out a series of calculations for the energies of some important low-lying states of $^{229}$Th$^{3+}$ at different correlation levels including Dirac-Fock(DF), second-order many-body perturbation theory MBPT(2), LCCSD, and full CCSD calculations. We also classified the contributions from different terms, including the Zeroth-order DF Coulomb correction ${E^{(0)}}$, the Breit corrections, second-order MBPT correction $E^{(2)}$, linearized part of single ($S_{1}$) and double ($S_{2}$) excitations correction ${E^{(S_{1}+S_{2})}}$, the non-linear coupled-cluster terms corrections ${E^{\rm(NL)}}$, and the QED correction. All these results listed in Table~\ref{energy1} comparing with experimental values~\cite{NIST,actinide} labeled as ${ E_{\rm Expt.}}$. The ${E_{\rm MBPT(2)}}$=${E^{(0)}}$+Breit+${ E^{(2)}}$ , ${ E_{\rm LCCSD}}$=${E^{(0)}}$+ Breit+${E^{(S_{1}+S_{2})}}$ , ${E_{\rm CCSD}}$=${ E_{\rm LCCSD}}$+ ${E^{\rm (NL)}}$, ${ E_{\rm Final}}$=${E_{\rm CCSD}}$+${\rm QED}$, represent the results obtained within second-order MBPT, LCCSD, CCSD, and CCSD results with QED correction approximations, respectively.

From the Table~\ref{energy1}, the importance of inclusion of correlation effect can be found. Take the incorrect determination of the ground state in the DF calculation as an example: the spectral measurements confirm that the ground state of Th$^{3+}$ is $5f_{5/2}$ state, while the ground state generated from DF calculation is the $6d_{3/2}$ state. The MBPT(2) results obviously overestimated the correlation effect and predicted much lower energies for all the states tabulated. The contribution from QED correction is about $0.1\%$ for $5f$ and $7s$ states, while it is negligible for other states. The CCSD results show a better agreement with experiments since the contribution from the nonlinear single- and double-excitation terms, which is about $0.3\%$ for $5f$ and $0.2\%$ for other states, has been included. Finally, for the two $5f$ states, the differences between ${\rm E_{Final}}$ and the experimental values are about $0.5\%$. The main reason for this discrepancy is the omission of the higher-order correlation effects, such as the triple- and higher excitation terms.

Table~\ref{energy2} presents a detailed comparison of our LCCSD, CCSD, and CCSD+QED transition energies with other $ab$ $initio$ results. It can be observed that in terms of transition energy, our CCSD results are consistent with the results obtained by relativistic intermediate Hamiltonian Fock-space coupled-cluster methods (FSCCSD)~\cite{ELIAV2012CP}, and both are much closer to the experimental value than the results obtained by the all-order calculations including partial extra third-order terms~\cite{Safronova2006PRA}. The contributions from different corrections are given in the all-order calculations~\cite{Safronova2006PRA}, allowing us to explore the origin of the difference with our results. It can be found that the DF Coulomb energies ${E^{(0)}}$ obtained from Ref.~\cite{Safronova2006PRA} and our calculations agree well with each other, while the absolute values of Breit correction to energies in Ref.~\cite{Safronova2006PRA} are significantly above our estimation. In addition, we also calculated the energy under the condition of $\ell_{max}=6$, and we found that our MBPT(2) and LCCSD results are completely consistent with their corresponding results. Comparing with the current calculation results of $\ell_{max}=7$, we observe that for the energies $5f$ states, the contributions of the partial wave $\ell=7$ are about $-$520 ${\rm cm^{-1}}$. This suggests that the contribution of higher-order fractional waves to energy is important, and the same conclusion also can be drawn in the work of Ref.~\cite{Safronova2006PRA}.

\subsection{Magnetic dipole moment}

\begin{table*}[ht]
  \begin{threeparttable}
  \newcommand{\RNum}[1]{\uppercase\expandafter{\romannumeral #1\relax}}
  \caption{Determination of the $^{229}$Th nuclear magnetic dipole moment using measured $A$ (MHz) from Ref.~\cite{Campbell2011prl} and the calculated $A/\mu$ from the present work. The Present $A/\mu$ column contains \textit{ab initio }results at different correlation levels in MHz/$\mu_{N}$. The results deduced from SDpT calculation and other methods are also listed here for comparison. }\label{hfsa}
  \begin{ruledtabular}
  \begin {tabular}{cccccccccccc}
   \multicolumn{1}{c}{\multirow{2}{*}{Level}}
   &\multicolumn{1}{c}{\multirow{2}{*}{${{A_{\rm Expt.}}}$}}
   &\multicolumn{4}{c}{Present $A/\mu$}
   &\multicolumn{1}{c}{Other $A/\mu$}
   &\multicolumn{4}{c}{$\mu$}\\
  \cline{3-6}\cline{7-7}\cline{8-11}
  \multicolumn{1}{c}{}
  &\multicolumn{1}{c}{}
  &\multicolumn{1}{c}{DF}
  &\multicolumn{1}{c}{MBPT(3)}
  &\multicolumn{1}{c}{LCCSD}
  &\multicolumn{1}{c}{CCSD}
  &\multicolumn{1}{c}{SDpT~\cite{Safronova2013PRA2}}
  &\multicolumn{1}{c}{LCCSD}
  &\multicolumn{1}{c}{CCSD}
  &\multicolumn{1}{c}{SDpT~\cite{Safronova2013PRA2}}
  &\multicolumn{1}{c}{Others}\\
  \hline
    $5f_{5/2}$&82.2(6)   &203.73 &250.52   &232.28     &230.53   &229.2   &0.354(3)  & 0.357(3)   &	0.359   &\\
    $5f_{7/2}$&31.4(7)   &105.79 &81.49    &87.14      &87.12    &86.1    &0.360(8)  & 0.360(8)   &	0.365   &\\
    $6d_{3/2}$&155.3(12) &331.75 &455.21   &442.71     &431.85   &431.5   &0.351(3)  & 0.360(3)   &	0.360   &\\
    $6d_{5/2}$&$-$12.6(7)&121.73 &$-$34.43 &$-$35.39   &$-$23.31 &$-$36.7 &0.356(20) & 0.541(30)  &	0.343   &\\
    Final     &          &       &         &           &         &        &0.355(9)  & 0.359(9)   & 0.360(7)&0.530$-$0.655~\cite{Minkov2019PRL}\\
              &          &       &         &           &         &        &          &            &         &0.54~\cite{Chasman1977RMP}\\
              &          &       &         &           &         &        &          &            &         &0.46(4)~\cite{Gerstenkorn1974JPF}\\
  \end{tabular}
  \end{ruledtabular}
  \end{threeparttable}
  \end{table*}

  \begin{table*}[ht]
    \begin{threeparttable}
    \newcommand{\RNum}[1]{\uppercase\expandafter{\romannumeral #1\relax}}
    \caption{Determination of the $^{229}$Th nuclear electric quadrupole moment using measured $B$ (MHz) from Ref.~\cite{Campbell2011prl} and the calculated $B/Q$ from the present work. The Present $B/Q$ column contains\textit{ ab initio} results at different correlation levels in MHz/eb.. The results deduced from SDpT calculation are also listed.}\label{hfsb}
    \begin{ruledtabular}
    \begin {tabular}{cccccccccc}
     \multicolumn{1}{c}{\multirow{2}{*}{Level}}
     &\multicolumn{1}{c}{\multirow{2}{*}{${{B_{\rm Expt.}}}$}}
     &\multicolumn{4}{c}{Present $B/Q$}
     &\multicolumn{1}{c}{Other $B/Q$}
     &\multicolumn{3}{c}{$Q$}\\
    \cline{3-6}\cline{7-7}\cline{8-10}

    \multicolumn{1}{c}{}
    &\multicolumn{1}{c}{}
    &\multicolumn{1}{c}{DF}
    &\multicolumn{1}{c}{MBPT(3)}
    &\multicolumn{1}{c}{LCCSD}
    &\multicolumn{1}{c}{CCSD}
    &\multicolumn{1}{c}{SDpT~\cite{Safronova2013PRA2}}
    &\multicolumn{1}{c}{LCCSD}
    &\multicolumn{1}{c}{CCSD}
    &\multicolumn{1}{c}{SDpT~\cite{Safronova2013PRA2}}\\

    \hline
      $5f_{5/2}$&2269(6) &537 &436   &743  &783    & 725 &3.06(1)& 2.90(1) & 3.13 \\
      $5f_{7/2}$&2550(12)&575 &388   &829  &879    & 809 &3.07(2)& 2.90(2) & 3.15 \\
      $6d_{3/2}$&2265(9) &609 &717   &746  &758    & 738 &3.04(1)& 3.00(1) & 3.07 \\
      $6d_{5/2}$&2694(7) &648 &474   &883  &897    & 873 &3.05(1)& 3.00(1) & 3.09 \\
      Final     &        &    &      &     &       &     &3.05(3)& 2.95(7) &3.11(6)\\
    \end{tabular}
    \end{ruledtabular}
    \end{threeparttable}
    \end{table*}

The $A/\mu$ for the first four states of $^{229}$Th$^{3+}$ are calculated at different correlation levels including DF, third-order many-body perturbation theory MBPT(3), LCCSD and CCSD. The magnetic dipole moment $\mu$ are then extracted by combining our theoretical results with the experimental HFS constants~\cite{Campbell2011prl}. The results are listed in Table~\ref{hfsa}, and compared with other available calculated values~\cite{Gerstenkorn1974JPF,Chasman1977RMP,Minkov2019PRL,Safronova2013PRA2}. Uncertainties are given in parentheses. \\

As seen from Table~\ref{hfsa}, the MBPT(3) calculations significantly overestimated the correlation effect compared with our CCSD results, which is consistent with the case of energies. For the $5f$ states, the LCCSD results are very close to the CCSD results, where the contribution of the nonlinear terms to $A/\mu$ is about $1\%$, while it is about $2.5\%$ for the $6d_{3/2}$ state. The main source of theoretical uncertainty is the electron correlation. States with smaller correlation effects tend to be more conducive to accurate calculation. The correlation effects ${\rm (CCSD-DF)/CCSD \times 100\%}$ and the non-linear effects ${\rm (CCSD-LCCSD)/CCSD \times 100\%}$ accounted to the $A/\mu$ of above states are no more than $25\%$ and $3\%$, respectively, with the only exception of $6d_{5/2}$ being $565\%$ and $-35\%$, respectively. Therefore, even if these calculations are carried out under the same theoretical framework, it is not guaranteed that the results of different states can achieve the same precision, because different states are different in sensitivity to various correlation effects. In this case, averaging the theoretical calculation results of several states can effectively reduce the uncertainty caused by different sensitivities of different states to the correlation effect. Here, we abandon the $6d_{5/2}$ state which is strongly dependent on the correlation effect, and take the average value of the other three states to obtain the final $\mu$ value, $i.e.$ 0.359(9), where the uncertainty in parentheses is entirely from the measurement.\\

Our CCSD value of $0.359(9)$ matches well with the all-order SDpT results $0.360(7)$, but is much lower than all other reported results~\cite{Minkov2019PRL,Chasman1977RMP,Gerstenkorn1974JPF}. For example, our results are about $50\%-80\%$ lower than the previous nuclear theory predictions, and about $30\%$ lower than the experimentally deduced values from Fourier spectroscopy in 1974~\cite{Gerstenkorn1974JPF}. It must be pointed out that the results obtained by combining spectroscopy measurements with atomic structure calculations are more reliable than the predictions directly using nuclear model theory. Furthermore, current laser spectroscopy measurement techniques and theoretical calculation method of atomic structure have improved lot over time, which is also one of the important reasons why we think the current results as well as Safronova \textit{et al.}'s results are more reliable than the 1974 results~\cite{Gerstenkorn1974JPF}. Remind that the all-order SDpT calculation corresponds to a full consideration of linear single- and double-excitation terms and further inclusion of some triple-excitations perturbatively, while our CCSD calculation includes all the linear and non-linear single- and double- excitation terms. Thus it is reasonable to extract the contribution of the triple-excitation terms from the comparison between SDpT results and our LCCSD results. The comparison of our CCSD result and SDpT result indicates that the contribution of the partial triple excitations and the non-linear term on $A/\mu$ to the first three states are really small. Another interesting feature we observed is that the $A/\mu$ of $6d_{5/2}$ state obtained by the SDpT method is $-36.7$, which is almost the same as our LCCSD value of $-35.4$. This is an independent validation for both the SDpT results and our LCCSD results, and this also suggests a strong dependence for the $A/\mu$ of $6d_{5/2}$ on the nonlinear single- and double-excitation terms.

\subsection{Electric quadrupole moments}
Similar with the case of magnetic dipole moment, we also determine the $^{229}$Th nuclear electric quadrupole moment $Q$ using measured $B$ (MHz) from Ref.~\cite{Campbell2011prl} and our calculated $B/Q$. The results are listed in Table~\ref{hfsb}, and compared with SDpT results~\cite{Safronova2013PRA2}. From Table~\ref{hfsb}, one can see that the correlation effects in the MBPT(3) calculation are significantly different from the LCCSD and full CCSD results. Except for the $6d_{3/2}$ state, the contributions from the third-order MBPT correction to the zeroth-order DF values are negative, while the corrections of linear terms (${\rm LCCSD-DF}$) and nonlinear coupled-cluster terms (${\rm CCSD-LCCSD}$) are positive for these four states, respectively. The total correlation effects of the $5f_{5/2,7/2}$ states are very strong, about 30\% of the total CCSD results, while the correlation effects of $6d_{3/2,5/2}$ states are relatively small. In addition, it can be found that the higher the angular momentum, the greater the total correlation effect. Furthermore, the contribution of non-linear terms accounts for approximately a quarter of the total correlation effect in $5f$ states, but less than a tenth in $6d$ states. Both the total correlation effects and the non-linear terms corrections are significantly larger than those in $A/\mu$ for each state, which is opposite of Ra$^{+}$~\cite{li2021}. Such strong and unusual correlation effects indicate that it is more difficult to accurately calculate the $B/Q$ of ground state of $^{229}$Th$^{3+}$. Here we take the average value of all four states as final $Q$ value, $i.e.$ 2.95(7), where the uncertainty in parentheses is also entirely from the measurement.

It is found that the $Q$ value from the SDpT calculation is significantly above the present final CCSD value. As we have stated, the contribution of the triple-excitation terms can be extracted from the comparison between the SDpT results and our LCCSD results, and the difference between LCCSD and CCSD results owes to the non-linear terms. Taking the present LCCSD result $3.05(3)$ as a reference, the SDpT result $3.11(6)$ and the CCSD result $2.95(7)$ are on both sides of it. It indicates that the contribution of partial triple-excitation terms is positive, about $2\%$, while the contribution of SD non-linear terms is negative, about $3\%$. This trend will lead to a cancellation. Therefore, in order to extract $Q$ accurately, it is important not only to evaluate the contribution of nonlinear terms accurately, but also to consider the contribution of triple-excitation terms completely. Furthermore, comprehensive observations show that the contribution of the nonlinear term to the $A/\mu$ is positive, while to the $B/Q$ it is negative. The relative ratio of the correlation represented by nonlinear terms ${\rm (CCSD-LCCSD)/(CCSD-DF)}$ in the $B/Q$ is much large than in the $A/\mu$, which is different from that of Fr and Ra$^{+}$.

\subsection{Magnetic octupole hyperfine interaction}
Recently, the rapid development of spectroscopy technology makes the extraction of some high-order nuclear moments possible. For example, the nuclear magnetic octupole moments $\Omega$ of $^{133}$Cs~\cite{Gerginov2003prl}, $^{137}$Ba$^{+}$~\cite{Lewty2013pra}, and $^{171}$Yb~\cite{Singh2013pra,Groote2021PRA} have been successfully determined. The $^{229}$Th$^{3+}$ is also a good candidate for the studies of nuclear structure beyond the first two electromagnetic moments, since its ground state $5f_{5/2}$ has a large angular momentum and thus a large coupling to $\Omega$. Safronova \textit{et al.} have suggested that this coupling effect may induce a hyperfine interval at a level of a few Hz~\cite{Safronova2013PRA2}. To extract $\Omega$, an accurate calculation of relevant octupole hyperfine interaction matrix elements, including diagonal and important non-diagonal matrix elements of low-lying states in $^{229}$Th$^{3+}$, is necessary. Once the accurate HFS constant $C$ is measured, these matrix elements can be used to determine the magnetic octupole moment immediately. In addition, the present octupole matrix elements are expected to possess a higher level of accuracy than those from the previous calculations using the third-order perturbation method mentioned in Refs.~\cite{Beloy2014PRL,Johnson1987PRA}. Then the excitation energy of the $^{229}$Th nucleus may be determined more accurately to a certain extent using the method reported by Beloy~\cite{Beloy2014PRL}.

Table~\ref{hfsc} listed the electronic octupole hyperfine interaction matrix elements from DF, LCCSD, CCSD calculations, including diagonal $C/\Omega$ in KHz/($\mu_{N}$$\times$b) and non-diagonal matrix elements in MHz of low-lying states. The uncertainty of these octupole parameters mainly comes from the unconsidered high-order correlation effects beyond CCSD, which are generally not greater than the contribution of the SD nonlinear terms ($i.e.$CCSD-LCCSD). Therefore, we take CCSD results as the recommended values listed in the ``Final'' column and the corresponding absolute value of the difference between CCSD and LCCSD results as the uncertainty enclosed in parentheses.
It can be observed from the Table~\ref{hfsc} that for $5f_{5/2,7/2}$ and $6d_{5/2}$ states, the correlation effects are significant compared with the relatively small matrix elements, which would lead to large uncertainties in the calculations. For the $6d_{3/2}$ state, the conclusion is opposite. It is worth noting that the contributions of the non-linear term on $C/\Omega$ to the $5f$ and $6d$ states are very large which are somewhat different from $A/\mu$. This may be owing to fact that the value of $C/\Omega$ is too small to be accurate enough. The larger HFS constant $C$ is, the more conducive to accurate calculation and measurement. Therefore, the results of $6d_{3/2}$ state are more suitable for roughly predicting the magnitude of HFS constant $C$. These matrix elements are of great significance to extract the $\Omega$ of $^{229}$Th nucleus.

\begin{table}[H]
  \begin{threeparttable}
  \newcommand{\RNum}[1]{\uppercase\expandafter{\romannumeral #1\relax}}
  \caption{ $C/\Omega$ in KHz/($\mu_{N}$$\times$b) and off-diagonal matrix elements in MHz from DF, LCCSD, CCSD calculations. The CCSD results are taken as the recommended values listed in the ``Final'' column and the corresponding absolute value of the difference between CCSD and LCCSD results as the uncertainty enclosed in parentheses.}\label{hfsc}
  \begin{ruledtabular}
  \begin {tabular}{ccccc}
  \toprule
    Level&${\rm DF}$&${\rm LCCSD}$ & ${\rm CCSD}$&Final
    \\   \hline
    $5f_{5/2}$	&	0.68 	&$-$0.60&$-$0.38  &$-$0.38(22)\\
    $5f_{7/2}$	&	0.31 	&	1.83 	&	1.12 	  &1.12(71)   \\
    $6d_{3/2}$	&	6.77 	&	7.91 	&	7.88 	  &7.88(3)    \\
    $6d_{5/2}$	&	1.84 	&	0.21 	&	0.05 	  &0.05(16)   \\
    \multicolumn{5}{c}{$\text { Off-diagonal matrix elements}$}\\
    $\langle 5f_{5/2}||{T}^{e}_{1}|| 5f_{7/2}\rangle$ &329  &1121  &1189 &1189(68)  \\
    $\langle 5f_{5/2}||{T}^{e}_{2}|| 5f_{7/2}\rangle$ &423  &605   &637  &637(32)   \\
    $\langle 6d_{3/2}||{T}^{e}_{1}|| 6d_{5/2}\rangle$ &355  &4180  &3962 &3962(218)  \\
    $\langle 6d_{3/2}||{T}^{e}_{2}|| 6d_{5/2}\rangle$ &695  &815   &830  &830(15)  \\

  \end{tabular}
  \end{ruledtabular}
  \end{threeparttable}
  \end{table}

\section{Conclusion}

In summary, we carried out a comprehensive study of the energies and HFS constants for several low-lying states in the $^{229}$Th$^{3+}$, using a relativistic CCSD method based on a large Gaussian basis set. To clarify the role of different correlation corrections, we also provide some intermediate results including the results from DF, MBPT, and LCCSD calculations. The Breit and QED corrections to the energies are also investigated. Our calculations indicate that more higher-order corrections not included in our calculations are important to further reduce the difference between the theoretical prediction and experimental measurement. The magnetic dipole moment $\mu$ and the electric quadrupole moment $Q$ of $^{229}$Th are obtained by combining our atomic calculation with the measured HFS $A$ and $B$ values. Our recommended $\mu$ value of 0.359(9) is in excellent agreement with the SDpT result of $0.360(7)$. The present electric quadrupole moment, $Q=2.95(7)$, is smaller than their recommended value, $Q=3.11(6)$, by about $5\%$, possibly owing to the fact that the electron correlations represented by the non-linear terms, omitted in SDpT calculations, contribute significantly to the electric quadrupole moment and reduce the values. Further analysis shows that the third-order many-body perturbation theory is not a very effective tool to generate sufficiently accurate nuclear moments $\mu$ and $Q$ of $^{229}$Th. It always strongly overestimates the magnitude and sometimes gives an incorrect sign of the correlation effect. Additionally, we also present the magnetic octupole HFS constants $C/\Omega$ and some important non-diagonal hyperfine transition matrix elements, which are required for further extracting the magnetic octupole moment $\Omega$ of $^{229}$Th nucleus. Our calculations also show that the $6d_{3/2}$ is a suitable state to carry out precise measurements of the hyperfine splittings from which $\Omega$ of $^{229}$Th can be inferred.

\begin{acknowledgments}
 We are grateful to Yong-jun Cheng of Shanxi Normal University and Ting-xian Zhang of Lanzhou University of Technology for many useful discussions.
 The work was supported by the National Natural Science Foundation of China (No. 115074094, No. 12074295, No. 11774386), the Strategic Priority Research Program of the Chinese Academy of Sciences Grant (No. XDB21030300), the Post-doctoral research project of SZTU (No. 202028555301011), and the project of Educational Commission of Guangdong Province of China (No.2020KTSCX124).

\end{acknowledgments}
%%%%%%%%%%%%%%%%%%%%%%%% begin thebibliography  %%%%%%%%%%%%%%%%%%%%%%%%%%
%\bibliography{Th}

\begin{thebibliography}{33}%
  \makeatletter
  \providecommand \@ifxundefined [1]{%
   \@ifx{#1\undefined}
  }%
  \providecommand \@ifnum [1]{%
   \ifnum #1\expandafter \@firstoftwo
   \else \expandafter \@secondoftwo
   \fi
  }%
  \providecommand \@ifx [1]{%
   \ifx #1\expandafter \@firstoftwo
   \else \expandafter \@secondoftwo
   \fi
  }%
  \providecommand \natexlab [1]{#1}%
  \providecommand \enquote  [1]{``#1''}%
  \providecommand \bibnamefont  [1]{#1}%
  \providecommand \bibfnamefont [1]{#1}%
  \providecommand \citenamefont [1]{#1}%
  \providecommand \href@noop [0]{\@secondoftwo}%
  \providecommand \href [0]{\begingroup \@sanitize@url \@href}%
  \providecommand \@href[1]{\@@startlink{#1}\@@href}%
  \providecommand \@@href[1]{\endgroup#1\@@endlink}%
  \providecommand \@sanitize@url [0]{\catcode `\\12\catcode `\$12\catcode
    `\&12\catcode `\#12\catcode `\^12\catcode `\_12\catcode `\%12\relax}%
  \providecommand \@@startlink[1]{}%
  \providecommand \@@endlink[0]{}%
  \providecommand \url  [0]{\begingroup\@sanitize@url \@url }%
  \providecommand \@url [1]{\endgroup\@href {#1}{\urlprefix }}%
  \providecommand \urlprefix  [0]{URL }%
  \providecommand \Eprint [0]{\href }%
  \providecommand \doibase [0]{http://dx.doi.org/}%
  \providecommand \selectlanguage [0]{\@gobble}%
  \providecommand \bibinfo  [0]{\@secondoftwo}%
  \providecommand \bibfield  [0]{\@secondoftwo}%
  \providecommand \translation [1]{[#1]}%
  \providecommand \BibitemOpen [0]{}%
  \providecommand \bibitemStop [0]{}%
  \providecommand \bibitemNoStop [0]{.\EOS\space}%
  \providecommand \EOS [0]{\spacefactor3000\relax}%
  \providecommand \BibitemShut  [1]{\csname bibitem#1\endcsname}%
  \let\auto@bib@innerbib\@empty
  %</preamble>

  \bibitem [{\citenamefont {{Kroger}}\ and\ \citenamefont
    {{Reich}}(1976)}]{Kroger1976NuPhA}%
    \BibitemOpen
    \bibfield  {author} {\bibinfo {author} {\bibfnamefont {L.~A.}\ \bibnamefont
    {{Kroger}}}\ and\ \bibinfo {author} {\bibfnamefont {C.~W.}\ \bibnamefont
    {{Reich}}},\ }\bibfield  {title} {\enquote {\bibinfo {title} {{Features of
    the low-energy level scheme of $^{229}$Th as observed in the
    {\ensuremath{\alpha}}-decay of $^{233}$U}},}\ }\href {\doibase
    10.1016/0375-9474(76)90494-2} {\bibfield  {journal} {\bibinfo  {journal}
    {Nucl. Phys. A}\ }\textbf {\bibinfo {volume} {259}},\ \bibinfo {pages}
    {29--60} (\bibinfo {year} {1976})}\BibitemShut {NoStop}%
  \bibitem [{\citenamefont {{Beck}}\ \emph {et~al.}(2007)\citenamefont {{Beck}},
    \citenamefont {{Becker}}, \citenamefont {{Beiersdorfer}}, \citenamefont
    {{Brown}}, \citenamefont {{Moody}}, \citenamefont {{Wilhelmy}}, \citenamefont
    {{Porter}}, \citenamefont {{Kilbourne}},\ and\ \citenamefont
    {{Kelley}}}]{Beck2007PhRvL}%
    \BibitemOpen
    \bibfield  {author} {\bibinfo {author} {\bibfnamefont {B.~R.}\ \bibnamefont
    {{Beck}}}, \bibinfo {author} {\bibfnamefont {J.~A.}\ \bibnamefont
    {{Becker}}}, \bibinfo {author} {\bibfnamefont {P.}~\bibnamefont
    {{Beiersdorfer}}}, \bibinfo {author} {\bibfnamefont {G.~V.}\ \bibnamefont
    {{Brown}}}, \bibinfo {author} {\bibfnamefont {K.~J.}\ \bibnamefont
    {{Moody}}}, \bibinfo {author} {\bibfnamefont {J.~B.}\ \bibnamefont
    {{Wilhelmy}}}, \bibinfo {author} {\bibfnamefont {F.~S.}\ \bibnamefont
    {{Porter}}}, \bibinfo {author} {\bibfnamefont {C.~A.}\ \bibnamefont
    {{Kilbourne}}}, \ and\ \bibinfo {author} {\bibfnamefont {R.~L.}\ \bibnamefont
    {{Kelley}}},\ }\bibfield  {title} {\enquote {\bibinfo {title} {{Energy
    Splitting of the Ground-State Doublet in the Nucleus Th229}},}\ }\href
    {\doibase 10.1103/PhysRevLett.98.142501} {\bibfield  {journal} {\bibinfo
    {journal} {\prl}\ }\textbf {\bibinfo {volume} {98}},\ \bibinfo {eid} {142501}
    (\bibinfo {year} {2007})}\BibitemShut {NoStop}%
  \bibitem [{\citenamefont {{Peik}}\ and\ \citenamefont
    {{Tamm}}(2003)}]{Peik2003EPL}%
    \BibitemOpen
    \bibfield  {author} {\bibinfo {author} {\bibfnamefont {E.}~\bibnamefont
    {{Peik}}}\ and\ \bibinfo {author} {\bibfnamefont {Chr.}\ \bibnamefont
    {{Tamm}}},\ }\bibfield  {title} {\enquote {\bibinfo {title} {{Nuclear laser
    spectroscopy of the 3.5 eV transition in Th-229}},}\ }\href {\doibase
    10.1209/epl/i2003-00210-x} {\bibfield  {journal} {\bibinfo  {journal}
    {Europhys. Lett.}\ }\textbf {\bibinfo {volume} {61}},\ \bibinfo {pages}
    {181--186} (\bibinfo {year} {2003})}\BibitemShut {NoStop}%
  \bibitem [{\citenamefont {Campbell}\ \emph {et~al.}(2012)\citenamefont
    {Campbell}, \citenamefont {Radnaev}, \citenamefont {Kuzmich}, \citenamefont
    {Dzuba}, \citenamefont {Flambaum},\ and\ \citenamefont
    {Derevianko}}]{Campbell2012PRL}%
    \BibitemOpen
    \bibfield  {author} {\bibinfo {author} {\bibfnamefont {C.~J.}\ \bibnamefont
    {Campbell}}, \bibinfo {author} {\bibfnamefont {A.~G.}\ \bibnamefont
    {Radnaev}}, \bibinfo {author} {\bibfnamefont {A.}~\bibnamefont {Kuzmich}},
    \bibinfo {author} {\bibfnamefont {V.~A.}\ \bibnamefont {Dzuba}}, \bibinfo
    {author} {\bibfnamefont {V.~V.}\ \bibnamefont {Flambaum}}, \ and\ \bibinfo
    {author} {\bibfnamefont {A.}~\bibnamefont {Derevianko}},\ }\bibfield  {title}
    {\enquote {\bibinfo {title} {Single-ion nuclear clock for metrology at the
    19th decimal place},}\ }\href {\doibase 10.1103/PhysRevLett.108.120802}
    {\bibfield  {journal} {\bibinfo  {journal} {Phys. Rev. Lett.}\ }\textbf
    {\bibinfo {volume} {108}},\ \bibinfo {pages} {120802} (\bibinfo {year}
    {2012})}\BibitemShut {NoStop}%
  \bibitem [{\citenamefont {{Peik}}\ \emph {et~al.}(2021)\citenamefont {{Peik}},
    \citenamefont {{Schumm}}, \citenamefont {{Safronova}}, \citenamefont
    {{P{\'a}lffy}}, \citenamefont {{Weitenberg}},\ and\ \citenamefont
    {{Thirolf}}}]{Peik2021QS}%
    \BibitemOpen
    \bibfield  {author} {\bibinfo {author} {\bibfnamefont {E.}~\bibnamefont
    {{Peik}}}, \bibinfo {author} {\bibfnamefont {T.}~\bibnamefont {{Schumm}}},
    \bibinfo {author} {\bibfnamefont {M.~S.}\ \bibnamefont {{Safronova}}},
    \bibinfo {author} {\bibfnamefont {A.}~\bibnamefont {{P{\'a}lffy}}}, \bibinfo
    {author} {\bibfnamefont {J.}~\bibnamefont {{Weitenberg}}}, \ and\ \bibinfo
    {author} {\bibfnamefont {P.~G.}\ \bibnamefont {{Thirolf}}},\ }\bibfield
    {title} {\enquote {\bibinfo {title} {{Nuclear clocks for testing fundamental
    physics}},}\ }\href {\doibase 10.1088/2058-9565/abe9c2} {\bibfield  {journal}
    {\bibinfo  {journal} {Quantum Sci. Technol.}\ }\textbf {\bibinfo {volume}
    {6}},\ \bibinfo {eid} {034002} (\bibinfo {year} {2021})},\ \Eprint
    {http://arxiv.org/abs/2012.09304} {arXiv:2012.09304 [quant-ph]} \BibitemShut
    {NoStop}%
  \bibitem [{\citenamefont {Safronova}\ \emph {et~al.}(2018)\citenamefont
    {Safronova}, \citenamefont {Budker}, \citenamefont {DeMille}, \citenamefont
    {Kimball}, \citenamefont {Derevianko},\ and\ \citenamefont
    {Clark}}]{Safronova2018RMP}%
    \BibitemOpen
    \bibfield  {author} {\bibinfo {author} {\bibfnamefont {M.~S.}\ \bibnamefont
    {Safronova}}, \bibinfo {author} {\bibfnamefont {D.}~\bibnamefont {Budker}},
    \bibinfo {author} {\bibfnamefont {D.}~\bibnamefont {DeMille}}, \bibinfo
    {author} {\bibnamefont {Jackson}\ \bibfnamefont {DerekF}\ \bibnamefont {Kimball}}, \bibinfo
    {author} {\bibfnamefont {A.}~\bibnamefont {Derevianko}}, \ and\ \bibinfo
    {author} {\bibfnamefont {C.~W.}\ \bibnamefont {Clark}},\ }\bibfield
    {title} {\enquote {\bibinfo {title} {Search for new physics with atoms and
    molecules},}\ }\href {\doibase 10.1103/RevModPhys.90.025008} {\bibfield
    {journal} {\bibinfo  {journal} {Rev. Mod. Phys.}\ }\textbf {\bibinfo {volume}
    {90}},\ \bibinfo {pages} {025008} (\bibinfo {year} {2018})}\BibitemShut
    {NoStop}%
  \bibitem [{\citenamefont {Fadeev}\ \emph {et~al.}(2020)\citenamefont {Fadeev},
    \citenamefont {Berengut},\ and\ \citenamefont {Flambaum}}]{Flambaum2020PRA}%
    \BibitemOpen
    \bibfield  {author} {\bibinfo {author} {\bibfnamefont {Pavel}\ \bibnamefont
    {Fadeev}}, \bibinfo {author} {\bibfnamefont {Julian~C.}\ \bibnamefont
    {Berengut}}, \ and\ \bibinfo {author} {\bibfnamefont {Victor~V.}\
    \bibnamefont {Flambaum}},\ }\bibfield  {title} {\enquote {\bibinfo {title}
    {{Sensitivity of $^{229}\mathrm{Th}$ nuclear clock transition to variation of
    the fine-structure constant}},}\ }\href {\doibase
    10.1103/PhysRevA.102.052833} {\bibfield  {journal} {\bibinfo  {journal}
    {Phys. Rev. A}\ }\textbf {\bibinfo {volume} {102}},\ \bibinfo {pages}
    {052833} (\bibinfo {year} {2020})}\BibitemShut {NoStop}%
  \bibitem [{\citenamefont {Minkov}\ and\ \citenamefont
    {P\'alffy}(2019)}]{Minkov2019PRL}%
    \BibitemOpen
    \bibfield  {author} {\bibinfo {author} {\bibfnamefont {Nikolay}\ \bibnamefont
    {Minkov}}\ and\ \bibinfo {author} {\bibfnamefont {Adriana}\ \bibnamefont
    {P\'alffy}},\ }\bibfield  {title} {\enquote {\bibinfo {title} {Theoretical
    predictions for the magnetic dipole moment of $^{229m}\mathrm{Th}$},}\ }\href
    {\doibase 10.1103/PhysRevLett.122.162502} {\bibfield  {journal} {\bibinfo
    {journal} {Phys. Rev. Lett.}\ }\textbf {\bibinfo {volume} {122}},\ \bibinfo
    {pages} {162502} (\bibinfo {year} {2019})}\BibitemShut {NoStop}%
  \bibitem [{\citenamefont {Flambaum}(2006)}]{Flambaum2006PRL}%
    \BibitemOpen
    \bibfield  {author} {\bibinfo {author} {\bibfnamefont {V.~V.}\ \bibnamefont
    {Flambaum}},\ }\bibfield  {title} {\enquote {\bibinfo {title} {Enhanced
    effect of temporal variation of the fine structure constant and the strong
    interaction in $^{229}\mathrm{Th}$},}\ }\href {\doibase
    10.1103/PhysRevLett.97.092502} {\bibfield  {journal} {\bibinfo  {journal}
    {Phys. Rev. Lett.}\ }\textbf {\bibinfo {volume} {97}},\ \bibinfo {pages}
    {092502} (\bibinfo {year} {2006})}\BibitemShut {NoStop}%
  \bibitem [{\citenamefont {Berengut}\ \emph {et~al.}(2009)\citenamefont
    {Berengut}, \citenamefont {Dzuba}, \citenamefont {Flambaum},\ and\
    \citenamefont {Porsev}}]{Berengut2009PRL}%
    \BibitemOpen
    \bibfield  {author} {\bibinfo {author} {\bibfnamefont {J.~C.}\ \bibnamefont
    {Berengut}}, \bibinfo {author} {\bibfnamefont {V.~A.}\ \bibnamefont {Dzuba}},
    \bibinfo {author} {\bibfnamefont {V.~V.}\ \bibnamefont {Flambaum}}, \ and\
    \bibinfo {author} {\bibfnamefont {S.~G.}\ \bibnamefont {Porsev}},\ }\bibfield
     {title} {\enquote {\bibinfo {title} {Proposed experimental method to
    determine $\ensuremath{\alpha}$ sensitivity of splitting between ground and
    7.6 ev isomeric states in $^{229}\mathrm{Th}$},}\ }\href {\doibase
    10.1103/PhysRevLett.102.210801} {\bibfield  {journal} {\bibinfo  {journal}
    {Phys. Rev. Lett.}\ }\textbf {\bibinfo {volume} {102}},\ \bibinfo {pages}
    {210801} (\bibinfo {year} {2009})}\BibitemShut {NoStop}%
  \bibitem [{\citenamefont {{Chasman}}\ \emph {et~al.}(1977)\citenamefont
    {{Chasman}}, \citenamefont {{Ahmad}}, \citenamefont {{Friedman}},\ and\
    \citenamefont {{Erskine}}}]{Chasman1977RMP}%
    \BibitemOpen
    \bibfield  {author} {\bibinfo {author} {\bibfnamefont {R.~R.}\ \bibnamefont
    {{Chasman}}}, \bibinfo {author} {\bibfnamefont {I.}~\bibnamefont {{Ahmad}}},
    \bibinfo {author} {\bibfnamefont {A.~M.}\ \bibnamefont {{Friedman}}}, \ and\
    \bibinfo {author} {\bibfnamefont {J.~R.}\ \bibnamefont {{Erskine}}},\
    }\bibfield  {title} {\enquote {\bibinfo {title} {{Survey of single-particle
    states in the mass region $A>228$}},}\ }\href {\doibase
    10.1103/RevModPhys.49.833} {\bibfield  {journal} {\bibinfo  {journal} {Rev.
    Mod. Phys.}\ }\textbf {\bibinfo {volume} {49}},\ \bibinfo {pages} {833--891}
    (\bibinfo {year} {1977})}\BibitemShut {NoStop}%
  \bibitem [{\citenamefont {Gerstenkorn}\ \emph {et~al.}(1974)\citenamefont
    {Gerstenkorn}, \citenamefont {Luc}, \citenamefont {Verges}, \citenamefont
    {Englekemeir}, \citenamefont {Gindler},\ and\ \citenamefont
    {Tomkins}}]{Gerstenkorn1974JPF}%
    \BibitemOpen
    \bibfield  {author} {\bibinfo {author} {\bibfnamefont {S.}~\bibnamefont
    {Gerstenkorn}}, \bibinfo {author} {\bibfnamefont {P.}~\bibnamefont {Luc}},
    \bibinfo {author} {\bibfnamefont {J.}~\bibnamefont {Verges}}, \bibinfo
    {author} {\bibfnamefont {D.~W.}\ \bibnamefont {Englekemeir}}, \bibinfo
    {author} {\bibfnamefont {J.~E.}\ \bibnamefont {Gindler}}, \ and\ \bibinfo
    {author} {\bibfnamefont {F.~S.}\ \bibnamefont {Tomkins}},\ }\bibfield
    {title} {\enquote {\bibinfo {title} {{{\rm Structures hyperfines du spectre
    d$'$$\acute{e}$tincelle,moment magn$\acute{e}$tique et quadrupolaire de
     l$'$isotope 229 du thorium}}},}\ }\href {\doibase 10.1051/jphys:01974003506048300} 
     {\bibfield{journal} {\bibinfo  {journal} {J. Phys. France}\ }\textbf {\bibinfo {volume}
    {35}},\ \bibinfo {pages} {483--495} (\bibinfo {year} {1974})}\BibitemShut
    {NoStop}%
  \bibitem [{\citenamefont {{Bemis}}\ \emph {et~al.}(1988)\citenamefont
    {{Bemis}}, \citenamefont {{McGowan}}, \citenamefont {{Ford}}, \citenamefont
    {{Milner}}, \citenamefont {{Robinson}}, \citenamefont {{Stelson}},
    \citenamefont {{Leander}},\ and\ \citenamefont {{Reich}}}]{Bemis_1988}%
    \BibitemOpen
    \bibfield  {author} {\bibinfo {author} {\bibfnamefont {C.~E.}\ \bibnamefont
    {{Bemis}}}, \bibinfo {author} {\bibfnamefont {F.~K.}\ \bibnamefont
    {{McGowan}}}, \bibinfo {author} {\bibfnamefont {Jr.}\ \bibnamefont {{Ford}},
    \bibfnamefont {J.~L.~C.}}, \bibinfo {author} {\bibfnamefont {W.~T.}\
    \bibnamefont {{Milner}}}, \bibinfo {author} {\bibfnamefont {R.~L.}\
    \bibnamefont {{Robinson}}}, \bibinfo {author} {\bibfnamefont {P.~H.}\
    \bibnamefont {{Stelson}}}, \bibinfo {author} {\bibfnamefont {G.~A.}\
    \bibnamefont {{Leander}}}, \ and\ \bibinfo {author} {\bibfnamefont {C.~W.}\
    \bibnamefont {{Reich}}},\ }\bibfield  {title} {\enquote {\bibinfo {title}
    {{Coulomb excitation of states in $^{229}$Th}},}\ }\href {\doibase
    10.1088/0031-8949/38/5/004} {\bibfield  {journal} {\bibinfo  {journal} {phys.
    scr.}\ }\textbf {\bibinfo {volume} {38}},\ \bibinfo {pages} {657--663}
    (\bibinfo {year} {1988})}\BibitemShut {NoStop}%
  \bibitem [{\citenamefont {Campbell}\ \emph {et~al.}(2011)\citenamefont
    {Campbell}, \citenamefont {Radnaev},\ and\ \citenamefont
    {Kuzmich}}]{Campbell2011prl}%
    \BibitemOpen
    \bibfield  {author} {\bibinfo {author} {\bibfnamefont {C.~J.}\ \bibnamefont
    {Campbell}}, \bibinfo {author} {\bibfnamefont {A.~G.}\ \bibnamefont
    {Radnaev}}, \ and\ \bibinfo {author} {\bibfnamefont {A.}~\bibnamefont
    {Kuzmich}},\ }\bibfield  {title} {\enquote {\bibinfo {title} {Wigner crystals
    of $^{229}\mathrm{Th}$ for optical excitation of the nuclear isomer},}\
    }\href {\doibase 10.1103/PhysRevLett.106.223001} {\bibfield  {journal}
    {\bibinfo  {journal} {Phys. Rev. Lett.}\ }\textbf {\bibinfo {volume} {106}},\
    \bibinfo {pages} {223001} (\bibinfo {year} {2011})}\BibitemShut {NoStop}%
  \bibitem [{\citenamefont {Safronova}\ \emph {et~al.}(2013)\citenamefont
    {Safronova}, \citenamefont {Safronova}, \citenamefont {Radnaev},
    \citenamefont {Campbell},\ and\ \citenamefont {Kuzmich}}]{Safronova2013PRA2}%
    \BibitemOpen
    \bibfield  {author} {\bibinfo {author} {\bibfnamefont {M.~S.}\ \bibnamefont
    {Safronova}}, \bibinfo {author} {\bibfnamefont {U.~I.}\ \bibnamefont
    {Safronova}}, \bibinfo {author} {\bibfnamefont {A.~G.}\ \bibnamefont
    {Radnaev}}, \bibinfo {author} {\bibfnamefont {C.~J.}\ \bibnamefont
    {Campbell}}, \ and\ \bibinfo {author} {\bibfnamefont {A.}~\bibnamefont
    {Kuzmich}},\ }\bibfield  {title} {\enquote {\bibinfo {title} {Magnetic dipole
    and electric quadrupole moments of the ${}^{229}$th nucleus},}\ }\href
    {\doibase 10.1103/PhysRevA.88.060501(R)} {\bibfield  {journal} {\bibinfo
    {journal} {Phys. Rev. A}\ }\textbf {\bibinfo {volume} {88}},\ \bibinfo
    {pages} {060501(R)} (\bibinfo {year} {2013})}\BibitemShut {NoStop}%
  \bibitem [{\citenamefont {{Pyykk{\"o}}}(2018)}]{2018MolPh}%
    \BibitemOpen
    \bibfield  {author} {\bibinfo {author} {\bibfnamefont {Pekka}\ \bibnamefont
    {{Pyykk{\"o}}}},\ }\bibfield  {title} {\enquote {\bibinfo {title} {{Year-2017
    nuclear quadrupole moments}},}\ }\href {\doibase
    10.1080/00268976.2018.1426131} {\bibfield  {journal} {\bibinfo  {journal}
    {Mol. Phys.}\ }\textbf {\bibinfo {volume} {116}},\ \bibinfo {pages}
    {1328--1338} (\bibinfo {year} {2018})}\BibitemShut {NoStop}%
  \bibitem [{\citenamefont {Sahoo}\ \emph {et~al.}(2015)\citenamefont {Sahoo},
    \citenamefont {Nandy}, \citenamefont {Das},\ and\ \citenamefont
    {Sakemi}}]{Sahoo2015PRA}%
    \BibitemOpen
    \bibfield  {author} {\bibinfo {author} {\bibfnamefont {B.~K.}\ \bibnamefont
    {Sahoo}}, \bibinfo {author} {\bibfnamefont {D.~K.}\ \bibnamefont {Nandy}},
    \bibinfo {author} {\bibfnamefont {B.~P.}\ \bibnamefont {Das}}, \ and\
    \bibinfo {author} {\bibfnamefont {Y.}~\bibnamefont {Sakemi}},\ }\bibfield
    {title} {\enquote {\bibinfo {title} {Correlation trends in the hyperfine
    structures of $^{210,212}\mathrm{Fr}$},}\ }\href {\doibase
    10.1103/PhysRevA.91.042507} {\bibfield  {journal} {\bibinfo  {journal} {Phys.
    Rev. A}\ }\textbf {\bibinfo {volume} {91}},\ \bibinfo {pages} {042507}
    (\bibinfo {year} {2015})}\BibitemShut {NoStop}%
  \bibitem [{\citenamefont {{Li}}\ \emph
    {et~al.}(2021{\natexlab{a}})\citenamefont {{Li}}, \citenamefont {{Tang}},
    \citenamefont {{Qiao}},\ and\ \citenamefont {{Shi}}}]{li2021}%
    \BibitemOpen
    \bibfield  {author} {\bibinfo {author} {\bibfnamefont {Feichen}\ \bibnamefont
    {{Li}}}, \bibinfo {author} {\bibfnamefont {Yong-Bo}\ \bibnamefont {{Tang}}},
    \bibinfo {author} {\bibfnamefont {Hao-xue}\ \bibnamefont {{Qiao}}}, \ and\
    \bibinfo {author} {\bibfnamefont {Ting-Yun}\ \bibnamefont {{Shi}}},\
    }\bibfield  {title} {\enquote {\bibinfo {title} {{Ab initio calculations of
    the hyperfine structure of Ra$^{+}$ and evaluations of the electric
    quadrupole moment Q of the 209,211,221,223Ra nuclei}},}\ }\href {\doibase
    10.1088/1361-6455/abcdf0} {\bibfield  {journal} {\bibinfo  {journal} {J.
    Phys. B: At. Mol. Opt. Phys.}\ }\textbf {\bibinfo {volume} {54}},\ \bibinfo
    {eid} {145004} (\bibinfo {year} {2021}{\natexlab{a}})}\BibitemShut {NoStop}%
  \bibitem [{\citenamefont {Beloy}(2014)}]{Beloy2014PRL}%
    \BibitemOpen
    \bibfield  {author} {\bibinfo {author} {\bibfnamefont {K.}~\bibnamefont
    {Beloy}},\ }\bibfield  {title} {\enquote {\bibinfo {title} {Hyperfine
    structure in ${^{229g}\mathrm{Th}}^{3+}$ as a probe of the
    $^{229g}\mathrm{Th}\ensuremath{\rightarrow}^{229m}\mathrm{Th}$ nuclear
    excitation energy},}\ }\href {\doibase 10.1103/PhysRevLett.112.062503}
    {\bibfield  {journal} {\bibinfo  {journal} {Phys. Rev. Lett.}\ }\textbf
    {\bibinfo {volume} {112}},\ \bibinfo {pages} {062503} (\bibinfo {year}
    {2014})}\BibitemShut {NoStop}%
  \bibitem [{\citenamefont {Roberts}\ \emph {et~al.}(2013)\citenamefont
    {Roberts}, \citenamefont {Dzuba},\ and\ \citenamefont
    {Flambaum}}]{Roberts2013PRA}%
    \BibitemOpen
    \bibfield  {author} {\bibinfo {author} {\bibfnamefont {B.~M.}\ \bibnamefont
    {Roberts}}, \bibinfo {author} {\bibfnamefont {V.~A.}\ \bibnamefont {Dzuba}},
    \ and\ \bibinfo {author} {\bibfnamefont {V.~V.}\ \bibnamefont {Flambaum}},\
    }\bibfield  {title} {\enquote {\bibinfo {title} {Parity nonconservation in
    fr-like actinide and cs-like rare-earth-metal ions},}\ }\href {\doibase
    10.1103/PhysRevA.88.012510} {\bibfield  {journal} {\bibinfo  {journal} {Phys.
    Rev. A}\ }\textbf {\bibinfo {volume} {88}},\ \bibinfo {pages} {012510}
    (\bibinfo {year} {2013})}\BibitemShut {NoStop}%
  \bibitem [{\citenamefont {{Li}}\ \emph
    {et~al.}(2021{\natexlab{b}})\citenamefont {{Li}}, \citenamefont {{Ma}},\ and\
    \citenamefont {{Tang}}}]{li2021JPhB}%
    \BibitemOpen
    \bibfield  {author} {\bibinfo {author} {\bibfnamefont {Fang}\ \bibnamefont
    {{Li}}}, \bibinfo {author} {\bibfnamefont {Hong}\ \bibnamefont {{Ma}}}, \
    and\ \bibinfo {author} {\bibfnamefont {Yong-Bo}\ \bibnamefont {{Tang}}},\
    }\bibfield  {title} {\enquote {\bibinfo {title} {{Relativistic
    coupled-cluster calculation of hyperfine-structure constants of
    La$^{2+}$}},}\ }\href {\doibase 10.1088/1361-6455/abcdf0} {\bibfield
    {journal} {\bibinfo  {journal} {J. Phys. B: At. Mol. Opt. Phys.}\ }\textbf
    {\bibinfo {volume} {54}},\ \bibinfo {eid} {065003} (\bibinfo {year}
    {2021}{\natexlab{b}})}\BibitemShut {NoStop}%
    \bibitem [{\citenamefont {Schwartz}(1955)}]{Schwartz1955PR}%
    \BibitemOpen
    \bibfield  {author} {\bibinfo {author} {\bibfnamefont {C.}~\bibnamefont
    {Schwartz}},\ }\href@noop {} {\bibfield  {journal} {\bibinfo  {journal}
    {Phys. Rev.}\ }\textbf {\bibinfo {volume} {97}},\ \bibinfo {pages} {380}
    (\bibinfo {year} {1955})}\BibitemShut {NoStop}%
  \bibitem [{\citenamefont {Chaudhuri}\ \emph {et~al.}(1999)\citenamefont
    {Chaudhuri}, \citenamefont {Panda},\ and\ \citenamefont
    {Das}}]{Chaudhuri1999pra}%
    \BibitemOpen
    \bibfield  {author} {\bibinfo {author} {\bibfnamefont {Rajat~K.}\
    \bibnamefont {Chaudhuri}}, \bibinfo {author} {\bibfnamefont {Prafulla~K.}\
    \bibnamefont {Panda}}, \ and\ \bibinfo {author} {\bibfnamefont {B.~P.}\
    \bibnamefont {Das}},\ }\bibfield  {title} {\enquote {\bibinfo {title} {Hybrid
    approach to relativistic gaussian basis functions: Theory and
    applications},}\ }\href {\doibase 10.1103/PhysRevA.59.1187} {\bibfield
    {journal} {\bibinfo  {journal} {Phys. Rev. A}\ }\textbf {\bibinfo {volume}
    {59}},\ \bibinfo {pages} {1187--1196} (\bibinfo {year} {1999})}\BibitemShut{NoStop}%
  \bibitem [{\citenamefont {Flambaum}\ and\ \citenamefont
    {Ginges}(2005)}]{Flambaum2005pra}%
    \BibitemOpen
    \bibfield  {author} {\bibinfo {author} {\bibfnamefont {V.~V.}\ \bibnamefont
    {Flambaum}}\ and\ \bibinfo {author} {\bibfnamefont {J.~S.~M.}\ \bibnamefont
    {Ginges}},\ }\bibfield  {title} {\enquote {\bibinfo {title} {Radiative
    potential and calculations of qed radiative corrections to energy levels and
    electromagnetic amplitudes in many-electron atoms},}\ }\href {\doibase
    10.1103/PhysRevA.72.052115} {\bibfield  {journal} {\bibinfo  {journal} {Phys.
    Rev. A}\ }\textbf {\bibinfo {volume} {72}},\ \bibinfo {pages} {052115}
    (\bibinfo {year} {2005})}\BibitemShut {NoStop}%
  \bibitem [{\citenamefont {Kramida}\ \emph {et~al.}(2020)\citenamefont
    {Kramida}, \citenamefont {{Yu.~Ralchenko}}, \citenamefont {Reader},\ and\
    \citenamefont {{and NIST ASD Team}}}]{NIST}%
    \BibitemOpen
    \bibfield  {author} {\bibinfo {author} {\bibfnamefont {A.}~\bibnamefont
    {Kramida}}, \bibinfo {author} {\bibnamefont {{Yu.~Ralchenko}}}, \bibinfo
    {author} {\bibfnamefont {J.}~\bibnamefont {Reader}}, \ and\ \bibinfo {author}
    {\bibnamefont {{and NIST ASD Team}}},\ }\href@noop {} {}\bibinfo
    {howpublished} {{NIST Atomic Spectra Database (ver. 5.8), [Online].
    Available: {\tt{https://physics.nist.gov/asd}} [2021, August 15]. National
    Institute of Standards and Technology, Gaithersburg, MD.}} (\bibinfo {year}
    {2020})\BibitemShut {NoStop}%
    \bibitem [{\citenamefont {Blaise}\ and\ \citenamefont {Wyart}()}]{actinide}%
    \BibitemOpen
    \bibfield  {author} {\bibinfo {author} {\bibfnamefont {J.}~\bibnamefont
    {Blaise}}\ and\ \bibinfo {author} {\bibfnamefont {J.}~\bibnamefont {Wyart}},\
    }\href {http://www.lac.universite-paris-saclay.fr/Data/Database/} { {\bibinfo
    {title} {http://www.lac.universite-paris-saclay.fr/Data/Database/}}}\BibitemShut{NoStop}%
  \bibitem [{\citenamefont {{Eliav}}\ and\ \citenamefont
    {{Kaldor}}(2012)}]{ELIAV2012CP}%
    \BibitemOpen
    \bibfield  {author} {\bibinfo {author} {\bibfnamefont {Ephraim}\ \bibnamefont
    {{Eliav}}}\ and\ \bibinfo {author} {\bibfnamefont {Uzi}\ \bibnamefont
    {{Kaldor}}},\ }\bibfield  {title} {\enquote {\bibinfo {title} {{Transition
    energies of Rn- and Fr-like actinide ions by relativistic intermediate
    Hamiltonian Fock-space coupled-cluster methods}},}\ }\href {\doibase
    10.1016/j.chemphys.2011.10.019} {\bibfield  {journal} {\bibinfo  {journal}
    {Chem. Phys.}\ }\textbf {\bibinfo {volume} {392}},\ \bibinfo {pages} {78--82}
    (\bibinfo {year} {2012})}\BibitemShut {NoStop}%
  \bibitem [{\citenamefont {Safronova}\ \emph {et~al.}(2006)\citenamefont
    {Safronova}, \citenamefont {Johnson},\ and\ \citenamefont
    {Safronova}}]{Safronova2006PRA}%
    \BibitemOpen
    \bibfield  {author} {\bibinfo {author} {\bibfnamefont {U.~I.}\ \bibnamefont
    {Safronova}}, \bibinfo {author} {\bibfnamefont {W.~R.}\ \bibnamefont
    {Johnson}}, \ and\ \bibinfo {author} {\bibfnamefont {M.~S.}\ \bibnamefont
    {Safronova}},\ }\bibfield  {title} {\enquote {\bibinfo {title} {Excitation
    energies, polarizabilities, multipole transition rates, and lifetimes in th
    iv},}\ }\href {\doibase 10.1103/PhysRevA.74.042511} {\bibfield  {journal}
    {\bibinfo  {journal} {Phys. Rev. A}\ }\textbf {\bibinfo {volume} {74}},\
    \bibinfo {pages} {042511} (\bibinfo {year} {2006})}\BibitemShut {NoStop}%
  \bibitem [{\citenamefont {Gerginov}\ \emph {et~al.}(2003)\citenamefont
    {Gerginov}, \citenamefont {Derevianko},\ and\ \citenamefont
    {Tanner}}]{Gerginov2003prl}%
    \BibitemOpen
    \bibfield  {author} {\bibinfo {author} {\bibfnamefont {Vladislav}\
    \bibnamefont {Gerginov}}, \bibinfo {author} {\bibfnamefont {Andrei}\
    \bibnamefont {Derevianko}}, \ and\ \bibinfo {author} {\bibfnamefont
    {Carol~E.}\ \bibnamefont {Tanner}},\ }\bibfield  {title} {\enquote {\bibinfo
    {title} {Observation of the nuclear magnetic octupole moment of
    $^{133}\mathrm{C}\mathrm{s}$},}\ }\href {\doibase
    10.1103/PhysRevLett.91.072501} {\bibfield  {journal} {\bibinfo  {journal}
    {Phys. Rev. Lett.}\ }\textbf {\bibinfo {volume} {91}},\ \bibinfo {pages}
    {072501} (\bibinfo {year} {2003})}\BibitemShut {NoStop}%
  \bibitem [{\citenamefont {Lewty}\ \emph {et~al.}(2013)\citenamefont {Lewty},
    \citenamefont {Chuah}, \citenamefont {Cazan}, \citenamefont {Barrett},\ and\
    \citenamefont {Sahoo}}]{Lewty2013pra}%
    \BibitemOpen
    \bibfield  {author} {\bibinfo {author} {\bibfnamefont {Nicholas~C.}\
    \bibnamefont {Lewty}}, \bibinfo {author} {\bibfnamefont {Boon~Leng}\
    \bibnamefont {Chuah}}, \bibinfo {author} {\bibfnamefont {Radu}\ \bibnamefont
    {Cazan}}, \bibinfo {author} {\bibfnamefont {Murray~D.}\ \bibnamefont
    {Barrett}}, \ and\ \bibinfo {author} {\bibfnamefont {B.~K.}\ \bibnamefont
    {Sahoo}},\ }\bibfield  {title} {\enquote {\bibinfo {title} {Experimental
    determination of the nuclear magnetic octupole moment of ${}^{137}$ba${}^{+}$
    ion},}\ }\href {\doibase 10.1103/PhysRevA.88.012518} {\bibfield  {journal}
    {\bibinfo  {journal} {Phys. Rev. A}\ }\textbf {\bibinfo {volume} {88}},\
    \bibinfo {pages} {012518} (\bibinfo {year} {2013})}\BibitemShut {NoStop}%
  \bibitem [{\citenamefont {Singh}\ \emph {et~al.}(2013)\citenamefont {Singh},
    \citenamefont {Angom},\ and\ \citenamefont {Natarajan}}]{Singh2013pra}%
    \BibitemOpen
    \bibfield  {author} {\bibinfo {author} {\bibfnamefont {Alok~K.}\ \bibnamefont
    {Singh}}, \bibinfo {author} {\bibfnamefont {D.}~\bibnamefont {Angom}}, \ and\
    \bibinfo {author} {\bibfnamefont {Vasant}\ \bibnamefont {Natarajan}},\
    }\bibfield  {title} {\enquote {\bibinfo {title} {Observation of the nuclear
    magnetic octupole moment of ${}^{173}$yb from precise measurements of the
    hyperfine structure in the ${{}^{3}P}_{2}$ state},}\ }\href {\doibase
    10.1103/PhysRevA.87.012512} {\bibfield  {journal} {\bibinfo  {journal} {Phys.
    Rev. A}\ }\textbf {\bibinfo {volume} {87}},\ \bibinfo {pages} {012512}
    (\bibinfo {year} {2013})}\BibitemShut {NoStop}%
  \bibitem [{\citenamefont {de~Groote}\ \emph {et~al.}(2021)\citenamefont
    {de~Groote}, \citenamefont {Kujanp\"a\"a}, \citenamefont {Koszor\'us},
    \citenamefont {Li},\ and\ \citenamefont {Moore}}]{Groote2021PRA}%
    \BibitemOpen
    \bibfield  {author} {\bibinfo {author} {\bibfnamefont {R.~P.}\ \bibnamefont
    {de~Groote}}, \bibinfo {author} {\bibfnamefont {S.}~\bibnamefont
    {Kujanp\"a\"a}}, \bibinfo {author} {\bibfnamefont {\'A.}\ \bibnamefont
    {Koszor\'us}}, \bibinfo {author} {\bibfnamefont {J.~G.}\ \bibnamefont {Li}},
    \ and\ \bibinfo {author} {\bibfnamefont {I.~D.}\ \bibnamefont {Moore}},\
    }\bibfield  {title} {\enquote {\bibinfo {title} {Magnetic octupole moment of
    $^{173}\mathrm{Yb}$ using collinear laser spectroscopy},}\ }\href {\doibase
    10.1103/PhysRevA.103.032826} {\bibfield  {journal} {\bibinfo  {journal}
    {Phys. Rev. A}\ }\textbf {\bibinfo {volume} {103}},\ \bibinfo {pages}
    {032826} (\bibinfo {year} {2021})}\BibitemShut {NoStop}%
  \bibitem [{\citenamefont {Johnson}\ \emph {et~al.}(1987)\citenamefont
    {Johnson}, \citenamefont {Idrees},\ and\ \citenamefont
    {Sapirstein}}]{Johnson1987PRA}%
    \BibitemOpen
    \bibfield  {author} {\bibinfo {author} {\bibfnamefont {W.~R.}\ \bibnamefont
    {Johnson}}, \bibinfo {author} {\bibfnamefont {M.}~\bibnamefont {Idrees}}, \
    and\ \bibinfo {author} {\bibfnamefont {J.}~\bibnamefont {Sapirstein}},\
    }\bibfield  {title} {\enquote {\bibinfo {title} {Second-order energies and
    third-order matrix elements of alkali-metal atoms},}\ }\href {\doibase
    10.1103/PhysRevA.35.3218} {\bibfield  {journal} {\bibinfo  {journal} {Phys.
    Rev. A}\ }\textbf {\bibinfo {volume} {35}},\ \bibinfo {pages} {3218--3226}
    (\bibinfo {year} {1987})}\BibitemShut {NoStop}%
  \end{thebibliography}

\end{document}